\documentclass[12pt]{article}
\usepackage{mathpazo}
\usepackage[T1]{fontenc}
\usepackage[utf8]{inputenc}

\usepackage{xcolor}
\usepackage{amsmath}
\usepackage{bm}
\usepackage{relsize}
\usepackage{comment}
\usepackage{graphicx}
\usepackage[UKenglish]{babel}
\usepackage{microtype}                      
\usepackage{xcolor} 
\usepackage{hyperref}                   	
\usepackage{lipsum}
\usepackage{fancyhdr}
\usepackage{picture} 
\hypersetup{								
	colorlinks=true,      		            
	linkcolor=blue,                        	
	filecolor=blue,                     	
	citecolor=blue,    	        			
	urlcolor=blue,                 	   		
	pdffitwindow=false,               	 	
	pdfstartview={FitH},               		
	pdfauthor={Huw Price},           		
}

\usepackage{braket}
\usepackage{physics}
\usepackage{todonotes}

\newcommand{\V}{$\lor$}

\newcommand{\upV}{$\wedge$}

\usepackage{tikz}
\usetikzlibrary{shapes,decorations,arrows,calc,arrows.meta,fit,positioning}
\tikzset{
    -Latex,auto,node distance =1 cm and 1 cm,semithick,
    state/.style ={ellipse, draw, minimum width = 0.7 cm},
    point/.style = {circle, draw, inner sep=0.04cm,fill,node contents={}},
    bidirected/.style={Latex-Latex,dashed},
    el/.style = {inner sep=2pt, align=left, sloped}
}

\begin{document}

\hypertarget{why-entanglement}{%
\title{Why Entanglement?}\label{why-entanglement}}
\author{Huw Price\thanks{University of Bonn and Trinity College, Cambridge, UK; email \href{mailto:hp331@cam.ac.uk}{hp331@cam.ac.uk}.} {\ and} Ken Wharton\thanks{Department of Physics and Astronomy, San Jos\'{e} State University, San Jos\'{e}, CA 95192-0106, USA; email \href{mailto:kenneth.wharton@sjsu.edu}{kenneth.wharton@sjsu.edu}.}}
\date{\today}
\maketitle\thispagestyle{empty}

\begin{abstract}
\noindent In this piece, written for a general audience, we propose a mechanism for quantum entanglement. The key ingredient is collider bias. In the language of causal models, a collider is a variable causally influenced by two or more other variables. Conditioning on a collider typically produces non-causal correlations between its contributing causes. This phenomenon can produce associations analogous to Bell correlations, in suitable post-selected ensembles. Such collider artefacts may become real connections, resembling causality, if a collider is `constrained' (e.g., by a future boundary condition). We consider the time-reversed analogues of these points in the context of retrocausal models of QM. Retrocausality yields a collider at the source of an EPR-Bell particle pair, and in this case constraint of the collider is possible by normal methods of experimental preparation. It follows that connections resembling causality may emerge across such colliders, from one branch of the experiment to the other. Our hypothesis is that this constrained retrocausal collider bias is the origin of entanglement. This piece is based on a suggestion first made in \cite{{PriceWharton21}}, and is an ancestor of an essay now published online in \textit{Aeon} magazine \cite{{PriceWharton23}}. In an updated version of the argument in \cite{PriceWharton23b} we (i) demonstrate its application in a real Bell experiment; and (ii) show that we can do without an explicit postulate of retrocausality.
\\

\noindent\textbf{Keywords:} Entanglement, collider bias, retrocausality
\end{abstract}

\section{Introduction}
Quantum entanglement has been in the news recently. The physicists Alain
Aspect, John Clauser and Anton Zeilinger won the 2022 Nobel Prize for,
as the citation puts it, `experiments with entangled photons,
establishing the violation of Bell inequalities and pioneering quantum
information science.'

Within the foundations of physics community, the prize has also been
welcomed as a tribute to John Stewart Bell (1928--1990),
the Irish
physicist who argued in the 1960s that quantum mechanics implies that
the world is unavoidably `non-local' -- that Einstein's `spooky action
at a distance' is unavoidable,
if quantum theory is correct. Clauser and
Aspect won their Nobel Prizes for pioneering work in the 1970s and 1980s
to test the relevant predictions of quantum mechanics. Such tests are
now called \emph{Bell experiments.}
Many increasingly sophisticated
versions, including Zeilinger's 'loophole-free' experiments, have confirmed that the world does
indeed behave as quantum theory predicts.

But \emph{why} does the world behave this way? Bell experiments rely on
quantum entanglement, which was first identified
by Erwin Schr\"odinger in 1935. He called it `not \emph{one} but rather
\emph{the} characteristic trait of quantum mechanics.' \cite{Sch35} 
More recently, it
has been described as `the essential fact of quantum mechanics' \cite{Suss14}, and `perhaps its weirdest feature' \cite{Weinberg13}. The
work of Clauser, Aspect, Zeilinger and others confirms the reality of
entanglement -- as Aspect himself put it, in his
\href{https://www.nobelprize.org/prizes/physics/2022/aspect/speech/}{{speech}}
at the Nobel Prize banquet, `entanglement is confirmed in its strangest
aspects' -- but it doesn't tell us what it is, or where it comes from.

Our \href{https://arxiv.org/abs/2101.05370}{{research}} \cite{{PriceWharton21}}
suggests a surprisingly simple answer. Entanglement may rest on a
familiar statistical phenomenon known as \emph{collider bias.} In the
light of collider bias, we think, entanglement is not really mysterious
at all. It is what we might have expected, if we'd taken seriously the
time-symmetry of the microworld.

Our proposal needs just three other ingredients, as well as collider
bias -- as we said, it is a simple recipe. All these ingredients are
available off the shelf (though admittedly, in one case, from a niche
corner of the shelf). But as far as we know, it has not been noticed
that they can be combined in this way, to throw new light on the central
puzzle of the quantum world.

\section{What is collider bias?}

Let's start with the main ingredient. Collider bias was described in the 1940s
by Joseph Berkson (1899--1982), a Mayo Clinic physicist, physician and
statistician.\footnote{Berkson was not the first to notice the point. It dates back at least to the Cambridge economist A C Pigou \cite{Pigou11}. (We are grateful to Jason Grossman and George Davey Smith here.)} Berkson noted an important source of error in
statistical reasoning used in medicine. In some circumstances, the
selection of a sample of patients produces misleading correlations
between their medical conditions. Taken at face value, these
correlations can suggest that one condition prevents another. Berkson
pointed out that these apparent causal connections may not be real. They
may be artefacts of the way the sample has been selected \cite{Berkson46}.

Simplifying Berkson's own example, imagine that all the patients
admitted to Ward C have similar symptoms, caused by one of two rare
infections, Virus A or Virus B. Ward C specialises in treating those
symptoms, so all its patients have at least one of these diseases. Some
may have both, but everyone on the ward who doesn't have Virus A is
certain to have Virus B, and vice versa.

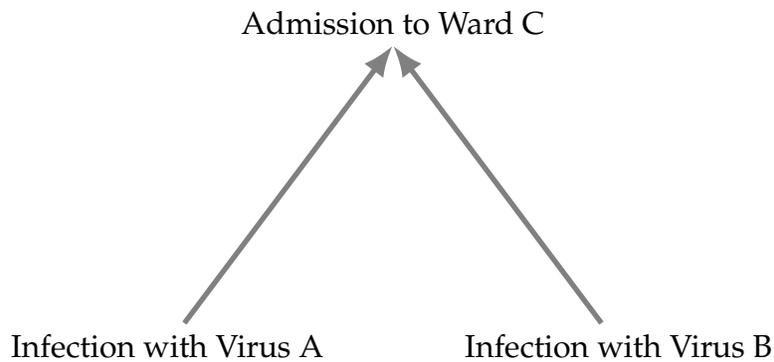
\begin{figure}
\centering
\begin{tikzpicture}
    \node (a) at (0,0) {Infection with Virus A};
    \node (F) at (3,4.3) {Admission to Ward C};
    \node (b) at (6,0) {Infection with Virus B};
\coordinate (coll) at (3,4); 
  \path[color=gray,rounded corners,line width=2pt] (a) edge (coll);
   \path[color=gray,rounded corners,line width=2pt] (b) edge (coll);

\end{tikzpicture}
\caption{A simple collider.} \label{fig:M1}
\end{figure}

Berkson's point was that this isn't evidence that avoiding one virus
leads to infection with the other one. The patients on Ward C are a very
biased sample. In the general population, having a vaccine for Virus A
won't make you more likely to catch Virus B.

This means that if a patient on Ward C with Virus A says to himself,
``I'm on Ward C, so if I hadn't caught Virus A I would have caught Virus
B'', then he's making a mistake. If he hadn't caught Virus A then (most
likely) he wouldn't have either virus, and he wouldn't have been
admitted to the ward.

This statistical effect is now called Berkson's bias, or \emph{collider
bias.} The term `collider' comes 
from causal modelling, the science of
inferring causes from statistical data. Causal modellers use diagrams
called Directed Acyclic Graphs (DAGs), made up of nodes linked by
arrows. The nodes represent events or states of affairs, and the arrows
represent causal connections between those events. When an event has two
independent contributing causes, like being a patient on Ward C, it is
shown in a DAG as a node where two arrows `collide' -- see Figure~\ref{fig:M1}.

If we just look at a sample of cases in which the event at a collider
happens, we'll often see a correlation between the two independent
causes. It may look like these causes are influencing one another, but
they are not. It is a selection artefact, as causal modellers say.
That's collider bias. The correlation stems from the way in which the
event at the collider depends on the two causes -- in our simple example
it needed one cause or the other.

\section{Rock-paper-scissors}

We want to take collider bias in the direction of physics -- ultimately,
in the direction of the kind of Bell experiments for which Clauser,
Aspect and Zeilinger won their Nobel Prizes. We want to propose an
explanation for what may be going on in those experiments, and other
cases of quantum entanglement.

We'll get there via a series of toy examples. For the first of them,
imagine that two physicists Alice and Bob play rock-paper-scissors,
sending their calls to a third observer, Charlie. Charlie makes a list
of the results: Alice wins, Bob wins, or it's a draw.

Suppose that Charlie likes Alice and dislikes Bob. He throws away most
of the results when Bob wins. In the remaining `official' results Alice
wins a lot more often than Bob. The correlation looks the way it would
if Alice actually had some influence over Bob's choice -- as though
Alice choosing scissors makes it a lot less likely that Bob will choose
rock, and so on. If Alice and Bob are far apart, this could look like
spooky action at a distance. But there's no real Alice-to-Bob causation
involved. It is just collider bias at work. The event at the collider --
whether Charlie retains or throws away the result -- is influenced both
by Alice's choice and by Bob's choice, giving us the same kind of
converging arrows as in Figure~\ref{fig:M1}.

Suppose that in a particular round of the game Alice chooses paper and
Bob chooses rock. As in the medical case, Alice would be making a
mistake if she says, ``If I had chosen scissors instead, Bob would
probably not have chosen rock.'' The right thing for her to say is, ``If
I had chosen scissors, then Charlie would probably have discarded the
result -- so my choice didn't make any difference to Bob's choice.''

\section{Constrained colliders}

Now to our second ingredient. It is the least familiar of all, though
it, too, is already on the shelf, if you know where to look. In the game
just described, Charlie could only favour Alice by discarding some
results. Let's see what happens if we rig the game in Alice's favour,
\emph{without throwing any results away.}

In our world this isn't going to happen naturally, so for now, let's
imagine it happening supernaturally. Suppose God also likes Alice more
than Bob, so he tweaks reality to give her an advantage. Perhaps he
arranges things so she never loses when she plays the game on Sundays.
How does God do it? It doesn't matter for our story, which doesn't need
to be realistic at this point, but here's one possibility. In a
deterministic universe everything that happens is determined by the
initial conditions at the very beginning of time. If God gets to choose
the initial conditions, and -- relying on a Laplacian calculation or
simply Divine foreknowledge -- knows exactly what follows from them,
then he can simply choose the initial conditions so that Alice never
loses on Sundays.

Readers who prefer a God-free version could imagine that Alice and Bob
live in a simulation, and that the superintelligence (AGI) that runs the simulation favours
Alice on Sundays. Some serious thinkers have suggested that we ourselves
may live in a simulation, so it would be hasty to say that this version
is inconceivable. But again, our example doesn't need to be realistic at
this point. Later in our argument, when realism matters, we won't have
to rely on God or simulations.

To invent some terminology, let's say that God (or the AGI)
\emph{constrains the collider} -- just on Sundays, in this version of
the story. To see what difference this makes, think again about a round
of the game where Alice chooses paper and Bob chooses rock. Is Alice
still making a mistake if she says, ``If I had chosen scissors instead,
Bob would not have chosen rock''? It now depends what day of the week it
is. This is still a mistake Monday through Saturday. On those days, the
right thing for Alice to say is, ``If I had chosen scissors, Bob would
still have chosen rock (and I would have lost).'' But Sunday is
different. On Sunday Alice never loses, so if she had chosen scissors,
Bob could not have chosen rock.

Let's suppose that Alice knows that the game works this way. Perhaps she
figured it out after years of experiments, and now makes a comfortable
living as a gambler, working one day a week. From her point of view, it
looks like she can \emph{control} Bob's choices, to some extent (and
only on Sundays). By choosing scissors she can \emph{prevent} Bob from
choosing rock, and so on.

With a \emph{constrained} collider, then, we would have something that
looks a lot like real causality across the collider, from one of the
pair of incoming causes to the other. True, it would be a very strange
kind of causality. For one thing, it would work the other way, too, from
Bob to Alice (though less happily, from his point of view). By choosing
rock on a Sunday Bob could prevent Alice from choosing scissors, and so
on.

For our purposes, it isn't going to matter whether this would be real
causality, or even whether the question makes sense, in this case. If we
press too hard on a toy example like this, it is liable to fall apart at
the seams. Could we still speak of both Alice and Bob as making free
choices, for example, if the choices are linked in this way?

All we need from the example is the following lesson. If there were
cases in nature in which something restricted the options at a collider,
we should expect to find a new kind of dependence between the normally
independent causes that feed into that collider. To keep our terminology
non-committal on the question whether it would count as causality, we'll
call this new kind of relation \emph{connection across a constrained
collider} (CCC).

CCC is our second ingredient, and certainly the least familiar one, for
most readers. But it, too, is already on the shelf, in the sense that
there's at least one place in physics where it has actually been
proposed. It is the key to a suggestion by Maldacena and Horowitz
\cite{HorowitzMaldacena04} for
solving the so-called black hole information paradox (made famous by
Stephen Hawking's
\href{https://en.wikipedia.org/wiki/Thorne%E2%80%93Hawking%E2%80%93Preskill_bet}{{bet}}
with Kip Thorne and John Preskill).

The Maldacena-Horowitz hypothesis relies on the proposal that special
`future boundary conditions' inside black holes constrain a collider (in
our terminology) at that point. Maldacena and Horowitz suggest that this
creates a zigzag causal path through time, along which information can
escape from a black hole.

Discussing the Maldacena-Horowitz hypothesis recently, the Cambridge
physicist Malcolm Perry says

\begin{quote}
{[}t{]}he interior of the black hole is therefore a strange place where
one's classical notions of causality \ldots{} are violated. This does
not matter as long as outside the black hole such pathologies do not
bother us. \cite[9]{Perry21}
\end{quote}
As we'll explain, our proposal is going to be that such pathologies are
actually extremely common, if you know where to look. In the other
direction of time, they are the basis of quantum entanglement -- and
they don't need black holes.

So far, then, we have two ingredients on the table: collider bias
itself, and CCC -- connection across a constrained collider. Before we
introduce the two remaining ingredients, let's get a bit closer to the
physics of the quantum world.

\section{From rock-paper-scissors to Bell experiments}

As we noted at the beginning, the recent Nobel Prizes were awarded for
Bell experiments. These confirmed the strange correlations, predicted by
quantum theory, which Bell took to show that the quantum world is
unavoidably non-local. Given that these so-called Bell correlations were
important enough to win Nobel Prizes, readers may be surprised to learn
that they can easily be reproduced in a version of our
rock-paper-scissors game. The only change we need is to have Alice and
Bob each flip a coin before they make their choice.

In this variant -- call it \emph{quantum} rock-paper-scissors
(QRPS) -- Alice and Bob each send two pieces of information to Charlie:
their choice of rock-paper-scissors, and the result of their coin flip.
So Charlie gets four values, two choices and two coin outcomes. This is
precisely the same amount of information generated in each run of a Bell
experiment. In that context, these values are called the two
\emph{measurement settings} and the two \emph{measurement outcomes.} The
Bell correlations are particular relationships between these four
values, in long lists of experimental results. (In one kind of Bell
experiment, for example, they specify among other things that whenever
the two settings are the same, the two outcomes must be different.)

In QRPS it is very easy for Charlie to set up a
filter, keeping some results and throwing away others, to make sure that
the set of results he keeps satisfies the Bell correlations. By using
the right filter, Charlie can ensure that the selected results look
\emph{exactly} like the data generated in the familiar kind of Bell
experiment of Clauser, Aspect and Zeilinger.\footnote{See \cite{Guido21} for the same point in the real QM context.}

Of course, this doesn't mean that there is any strange
non-locality in QRPS. As in the earlier version, the correlations are
a selection artefact, a result of collider bias. But since the
results are now a perfect copy of real Bell experiments, it is worth
paying careful attention to differences between the two.

\begin{figure}
\centering
\begin{tikzpicture}
    \node (a) at (0,0) {Alice};
    \node (F) at (2,4.3) {Charlie};
    \coordinate (v) at (2,4);
    \node (b) at (4,0) {Bob};

     \node (a2) at (5,4.3) {Alice};
    \node (F2) at (7,0) {Charlie};
    \coordinate (v2) at (7,0.3);
    \node (b2) at (9,4.3) {Bob};

   \path (a) edge[-] (v);
    \path (b) edge[-] (v);
     \path (a2) edge[-] (v2);
    \path (b2) edge[-] (v2);

     \coordinate (zero) at (-1,-1);
     \node (t) at (-1,5) {Time};
     \node (x) at (11,-1) {Space};

      \path (zero) edge (t);
      \path (zero) edge (x);

\end{tikzpicture}
\caption{\upV-shaped and \V-shaped experiments.} \label{fig:M2}
\end{figure}
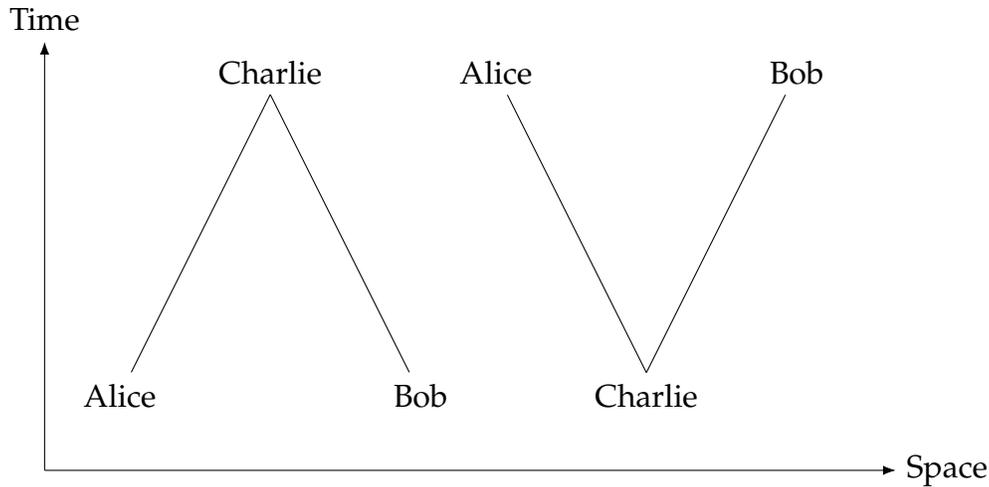

The most obvious difference is that quantum rock-paper-scissors and real
Bell experiments look like time-reversed versions of each other. In QRPS
Alice and Bob send their choices \emph{to} Charlie, later in time. In a
spacetime diagram with time running up the vertical axis, the structure
looks like an upside-down `V' -- see the left hand side of Figure~\ref{fig:M2}.
We'll say that cases like this are `\upV-shaped'. In real Bell experiments,
Alice and Bob receive their particles \emph{fro}m the source, which is
earlier in time. So the structure looks like \V, as in the right hand
side of Figure~\ref{fig:M2} -- we'll say that they are `\V-shaped'.

In a moment we're going to introduce a \V-shaped version of QRPS, to
eliminate this difference. But first let's summarise what we learned
from the \upV-shaped case. We saw that it is easy for Charlie to set up a
filter, to make sure that the results he keeps satisfy the Bell
correlations. But we don't need any non-locality between Alice and Bob,
to explain what's going on. The correlations are simply collider bias.
As in the original rock-paper-scissors case (without constraint at the
collider) they are simply a selection artefact.

We could reintroduce God or an AGI at this point, to add a constrained
collider to QRPS. There would be one interesting difference from the
original game. In that case, we saw that the effect of the constraint
was to give Alice and Bob control over each other's choices, making it
hard to maintain that they both had freedom to choose. In QRPS, as in
the analogous real Bell experiments, that problem goes away: Alice and
Bob each get some influence over the result of the other's coin toss,
but we can still treat both of their own choices as completely free.

That difference aside, a constrained version of QRPS would be as
unrealistic as for the original game, and wouldn't tell us anything new.
So let's turn instead to a \V-shaped version of QRPS, where realistic
constrained colliders will be much easier to find.

\section{Flipping the game}

As we said, QRPS looks like a time-reversed version of a Bell
experiment, \upV-shaped rather than \V-shaped. Instead of two particles
\emph{leaving} a common source and \emph{going to} separate observers,
it's the other way round. The information travels \emph{from} Alice and
Bob, \emph{to} Charlie at the common point in the future.

Can we flip this quantum rock-paper-scissors, to make it \V-shaped not
\upV-shaped? It might look easy. We can have Charlie toss the two coins and
send them to Alice and Bob, so that the results (heads or tails) become
Alice and Bob's measurement outcomes. But if that's all we do, Charlie
won't know what choices Alice and Bob are going to make when he sends
out the coins. That means there's no way for him to put bias into the
results, in the way that he could in the \upV-shaped case. There's no way
that Charlie can produce the Bell correlations, in other words.

But suppose we let Charlie \emph{know in advance} what choices Alice and
Bob are going to make -- we give him a crystal ball, say. Then it is
very easy for him to manage the coins so that the net results, gathered
over many plays of the game, satisfy the Bell correlations. The trick is
for Charlie to toss one coin, and then choose the result for the other
coin based on a rule that takes into account Alice and Bob's future
choices.

So this game, too, generates the same kind of Bell correlations as the
famous experiments of Clauser, Aspect, and Zeilinger. Let's ask the same
question we did about \upV-shaped QRPS. Does the new \V-shaped version
involve some kind of spooky action at a distance from Alice to Bob, and
vice versa?

\begin{figure}
\centering
\begin{tikzpicture}
    \node (a) at (0,0) {Alice's choice};
    \node (F) at (2.5,-4.3) {Charlie's selection of outcomes};
    \node (b) at (5,0) {Bob's choice};
\coordinate (coll) at (2.5,-4); 
  \path[color=gray,rounded corners,line width=2pt]  (a) edge (coll);
   \path[color=gray,rounded corners,line width=2pt] (b) edge (coll);

\end{tikzpicture}
\caption{A past collider.} \label{fig:M3}
\end{figure}
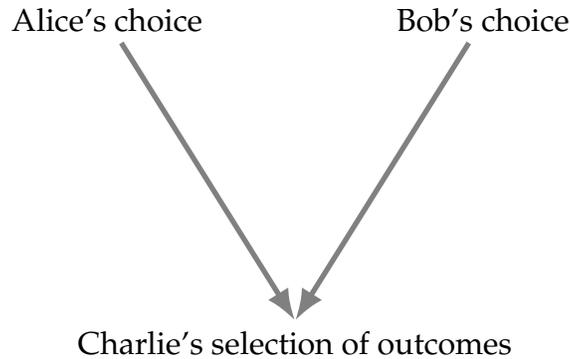

We hope that readers will be inclined to say `No' to this question.
After all, the basic causal structure of the new \V-shaped version is
something like Figure~\ref{fig:M3}. Thanks to Charlie's crystal ball, and the rule
he uses for selecting outcomes, Alice's and Bob's choices both influence
what outcomes Charlie produces, in every case. This means that Charlie's
selection procedure is a collider, and we have to be on our guard for
collider bias.

For this reason, attentive readers might suspect that collider bias
plays the same role in explaining the results the new \V-shaped QRPS as
it did in the \upV-shaped case. But there's one very big difference between
these two cases -- which brings us to our third ingredient.

\section{Initial control}

In the \upV-shaped version of QRPS, Charlie had to apply a filter, and
throw away results he didn't want. But in the \V-shaped case, he gets to
\emph{choose} the results, in the light of what he learns from the
crystal ball and his (single) coin toss. \emph{He doesn't have to throw
anything away.} In this case, then, Charlie himself can constrain the
collider. All he needs is an ordinary ability we take for granted, to
control the initial conditions of an experiment.

Perhaps we shouldn't take this for granted. It is actually a remarkable
ability, one that depends on the fact that we live in a place in which
abundant low-entropy energy can be harnessed by creatures like us. But
by ordinary standards, there's nothing surprising about it. We have much
more control over the \emph{initial} conditions of experiments than over
their \emph{final} conditions. It's easy to arrange the balls on a pool
table into precise positions \emph{before} the initial break, but
virtually impossible to play the game so that they all \emph{end up} in
those positions.

Let's call this familiar fact \emph{initial control.} It is the third
ingredient in our recipe.
Looking ahead a bit (so to speak), the final ingredient is going to be
something that does the job of the crystal balls, in giving Charlie
access to information about future choices by Alice and Bob. We're going
to find that on the shelf already in the quantum world, under the name
\emph{retrocausality.} But before we go there, let's summarise what we
learn from the new \V-shaped QRPS.

Thanks to the crystal balls, we have the causal structure shown in
Figure~\ref{fig:M3}, with a collider in the past, where Charlie chooses the
outputs. Thanks to initial control, it is easy to make it a constrained
collider, so that Charlie doesn't need to throw any results away. And
this means that there can be connection across the constrained collider
(CCC), from Alice to Bob, and vice versa.

\begin{figure}
\centering
\begin{tikzpicture}
    \node (a) at (0,0) {Alice's choice};
    \node (A) at (3,0) {Alice's outcome};
    \node (F) at (5,-4.3) {Charlie's selection of outcomes};
    \node (b) at (10,0) {Bob's choice};
    \node (B) at (7,0) {Bob's outcome};
\coordinate (coll) at (5,-4); 
\draw[color=blue,rounded corners,line width=2pt] (a) -- (5,-4) -- (B);
\draw[color=red,rounded corners,line width=2pt] (b) -- (5,-4) -- (A);
\end{tikzpicture}
\caption{The Parisian Zigzag.} \label{fig:M4}
\end{figure}
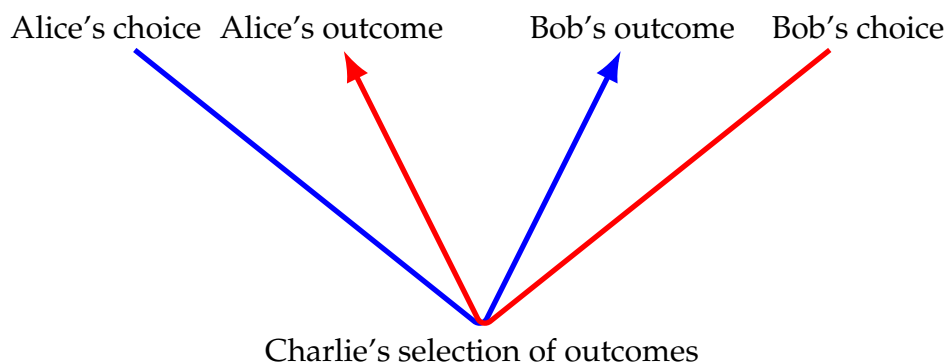

In other words, the combination of the collider structure in Figure~\ref{fig:M3}
and the constraint provided by initial control gives us CCC. If we are
happy to use causal language, we can say that it gives us the kind of
zigzag causal connections shown in Figure~\ref{fig:M4}, from Alice's choice to Bob's outcome and 
Bob's choice to Alice's outcome.\footnote{Two notes. First, there is disagreement about whether the term `causation' is appropriate in real Bell experiments, for the connection between setting choices on one side and outcomes on the other. For present purposes, we set aside that issue, though reasons for declining to use the term in real cases are likely to apply here, too, since the correlations are the same by definition. Second, we also set aside the question as to whether Alice and Bob should be said to influence their own outcomes; that is possible in real Bell experiments, of course, but doesn't provide a way to evade the argument for influence from each side to the other.} 
(Why do we call it a
Parisian Zigzag? More on that in a moment.)

In the light of this, let's go back to this question: does \V-shaped QRPS
involve some kind of spooky action at a distance from Alice to Bob, and
vice versa? At this point we need to be careful about what we mean by
action at a distance. As we have just seen, there is indeed some
influence, or connection, from Alice to Bob, and vice versa. Since they
are at a distance from each other, and a direct connection might need to
be faster than light, we might still want to call it action at a
distance, or nonlocality. (One of the recent Nobel laureates once told
us that he thought such a zigzag should still count as a nonlocal
effect.)

However, the connection between Alice and Bob is indirect, and depends
entirely on processes which don't themselves require anything faster
than light. So whatever we call it, it doesn't have the
relativity-challenging character normally associated with so-called
spooky action at a distance in QM. And it is not very mysterious: we
know exactly what it is, namely, connection across a constrained
collider.

The crystal balls were mysterious, of course, but once we gave
ourselves those, the explanation of the connection between Alice and Bob
is straightforward. Imagine if something like this could explain the
results of real Bell experiments -- that would be a nail in the coffin
of the quantum spooks!

\section{Retrocausality}

Well, let's see. We need our final ingredient. In \V-shaped QRPS, we gave
Charlie a crystal ball, to allow causation to work backwards -- in other
words, to allow Alice and Bob's choices to feed into the algorithm
Charlie uses to select the measurement outcomes. In the real world, of
course, we don't find magical crystal balls on any actual shelf.

In the quantum world, however, retrocausality is a familiar hypothesis.
In that sense, it is certainly available off the shelf. It was first
proposed in the late 1940s by the Parisian physicist Olivier Costa de
Beauregard \cite{Costa53}, who suggested that in the quantum world causal influence
might follow a zigzag path, as in Figure~\ref{fig:M4} -- that's why we called it
the Parisian Zigzag.

Retrocausality remained a niche idea for many years, though it has long
had some distinguished proponents. In the 1950s one of them, at least
briefly, was the British physicist Dennis Sciama \cite{Sciama58}, who taught an
astonishing generation of physicists, including Stephen Hawking. Sir
Roger Penrose, himself a recent Nobel laureate, has long been
sympathetic to the idea. There's a story from the 1990s of Penrose
drawing a zigzag at a quantum workshop at the Royal Society in London,
and joking `I can get away with proposing this kind of thing, because
I'm already a Fellow here.' (Now that he has a Nobel Prize it is even
easier, presumably!)\footnote{For a recent version of Penrose's views on this topic, see \cite{Penrose23}.} Another famous long-term proponent is the Israeli physicist Yakir Aharonov, whose interest dates from a well-known 1964 paper \cite{Aha64}.

More recently, we ourselves have written in
\href{https://aeon.co/essays/can-retrocausality-solve-the-puzzle-of-action-at-a-distance}{\emph{{Aeon}}}
and
\href{https://nautil.us/to-understand-your-past-look-to-your-future-235937/?_sp=e4739d6b-26e8-49bc-9403-92c611107706.1668570992307}{{elsewhere}}
about the advantages of retrocausal approaches to QM, both in avoiding
action at a distance, and in respecting time-symmetry. We argued that
quantum theory provides new reasons to think that microscopic
time-symmetry requires retrocausality \cite{PriceWharton16, WhartonPrice16}.

Recent popular discussions of the retrocausal approach to QM may be
found
\href{https://www.templeton.org/news/the-sudoku-universe}{{here}},
\href{https://www.newscientist.com/article/mg23731652-800-quantum-time-machine-how-the-future-can-change-what-happens-now/}{{here}} 
and
\href{https://phys.org/news/2017-07-physicists-retrocausal-quantum-theory-future.html}{{here}},
for example.\footnote{See \cite{Merali22, Becker18, Zyga17}. For specialist audiences, recent discussions include the following: \cite{PriceWharton15, LeiferPusey17, Leifer17, 
FriedrichEvans19, WhartonArgaman20, NorsenPrice21, Adlam22}.} It is now a sufficiently familiar proposal that no adequate
survey of the puzzles of quantum theory can afford to ignore it.
Some
commentators on the recent Nobel Prize announcement got themselves into
trouble 
for doing so \cite{Hance22}.

So retrocausality in QM is a well-known idea, and has well-known points
in its favour. However, an additional striking advantage seems to have
been overlooked. Retrocausality suggests a simple mechanism for
`\emph{the} characteristic trait of quantum mechanics' (Schr\"odinger),
`its weirdest feature' (Weinberg) -- in other words, for the strange
connections between separated systems called quantum entanglement.

Starting with retrocausality, our proposal goes like this, in four easy
steps. We've highlighted the use of our four ingredients.

\begin{enumerate}
\def\labelenumi{\arabic{enumi}.}
\item \textbf{Retrocausality} automatically introduces colliders into Bell
  experiments, at the point where the two particles are produced.

\item That's interesting because colliders produce \textbf{collider bias}
  and causal artefacts -- correlations that look like they involve
  causation, but really don't.
\item But \textbf{constraining a collider} can turn a causal artefact into a
  real connection across the collider.

\item In the case of colliders in the past, as in Figure 3, constraint is
  easy. It just follows from normal \textbf{initial control} of
  experiments.
\end{enumerate}

Taken together, these steps suggest a simple explanation for the
Parisian Zigzag, and the strange kind of non-local connections in the
quantum world, revealed by Bell's arguments. It is \emph{connection
across constrained colliders,} where the colliders result from
retrocausality and the constraints from ordinary initial control of
experimental setups.

Of course, more work is needed to show that this simple mechanism can
actually explain quantum entanglement. We don't mean to claim that it is
a trivial step from \V-shaped QRPS to real Bell experiments. What we do
claim, and what we take to be demonstrated by \V-shaped QRPS, is that the
combination of retrocausality and initial control can give rise to a
connection between separated systems that looks very similar to
entanglement. In our view, this is such a striking fact -- and
entanglement is otherwise such a strange and mysterious beast -- that we
propose the following hypothesis:

\begin{quote}
    \textbf{Hypothesis (E=CCC)}: Quantum entanglement is retrocausal collider bias,
constrained   by initial control.
\end{quote}

\noindent If this hypothesis turns out to be true, then in place of spooky action
at a distance, we'll get Costa de Beauregard's zigzag connections
-- which, as he always emphasised, are much easier to reconcile with
relativity -- but now explained as CCC.  It will still be true that quantum theory gives us a new
kind of connection between the properties of distant systems. The
experiments of Clauser, Aspect and Zeilinger provide very convincing
evidence that quantum entanglement is a real phenomenon. But it would no
longer look mysterious. On the contrary, any world that combines
retrocausality and initial control would be expected to look like this.

In earlier work, as we noted above,
\href{https://aeon.co/essays/can-retrocausality-solve-the-puzzle-of-action-at-a-distance}{{we}}
and
\href{https://phys.org/news/2017-07-physicists-retrocausal-quantum-theory-future.html}{{others}} 
have argued that time-symmetry requires retrocausality, once the world
becomes quantised \cite{PriceWharton16, Zyga17}.\footnote{See also \cite{Price12, PriceWharton15, LeiferPusey17, Leifer17}.} The same line of argument now seems to lead to
entanglement, once initial control is added to the picture. That's what
we meant at the beginning, when we said that entanglement is what we
might have expected, if we'd taken seriously the time-symmetry of the
microworld.\footnote{In our most recent version of this proposal \cite{PriceWharton23b}, we argue that at least for an operational version of the argument, the retrocausality does not need to be an independent postulate. In so far as it is needed, it comes for free when we consider an idealised regime without initial control.}

\section{Avoiding causal loops and signalling}

Finally, a note for readers who are worried that the cure is worse than
the disease -- that retrocausality opens the door to a menagerie of
paradoxes and problems. The note says: Well spotted! As we described
\V-shaped QRPS, with the crystal balls, you are absolutely right. For one
thing, the crystal balls give Charlie options much like those of the
famous time-traveller, encountering his own grandfather long before his
parents met. What's to stop Charlie interfering with the course of history,
say by bribing Bob to make a different choice than the one shown in the
crystal ball? (In the causal loop literature, this is called `bilking'.)

Also -- less dramatic, maybe, but especially interesting in comparison
to QM -- the crystal balls allow Alice and Bob to send messages to
Charlie, and hence potentially, with his help, to signal to each other.
This isn't possible in real Bell experiments, where Alice and Bob can't
signal to each other, despite having some influence on each other's
measurement outcomes. So isn't this bad news for retrocausality?

These are good objections, but it is easy to modify the \V-shaped QRPS
game to avoid them. What we need to do is to split Charlie's functions
into two parts. Most of what he does gets replaced by a simple
algorithm, inside a black box, that takes in information about the two
future measurement settings, and spits out the two measurement outcomes.
Charlie himself can't see inside the black box, and doesn't have access
to the future settings. But he still has a vital job to do. The box has
a knob on the front, with a small number of options. Charlie controls
that knob, and if he wants the device to produce the Bell correlations,
he needs to choose the right option. In the terminology of QM, that's
called `preparing the initial state'. If that's all that Charlie does,
and the quantum black box takes care of the rest, the door to the
menagerie is closed. Alice and Bob can no longer signal to Charlie, or
to each other. Everything works as in orthdox QM, except that we now
have the prospect of an explanation of entanglement.

This means that if nature wants retrocausality without retrosignalling
(and without the paradoxes that retrosignalling would lead to), it is
going to need black boxes -- places in nature where observers like
Charlie can't see the whole story. In normal circumstances, such black
boxes would seem like another kind of magic. Charlie is a clever guy,
after all. What's to stop him taking a peek inside any kind of box?

But in the quantum case, many readers will already know the answer to
this question. What's to stop Charlie taking a peek is Werner
Heisenberg, or more precisely his famous Uncertainty Principle. Ever
since Heisenberg, quantum theory has been built on the idea that there
are new limits to what it is possible to know about physical reality.
One of the central questions is whether this is just a restriction on
our knowledge of reality, or whether reality itself is somehow fuzzy. As
Schr\"odinger put it in 1935, after describing his famous Cat Experiment:
`There is a difference between a shaky or out-of-focus photograph and a
snapshot of clouds and fog banks.' \cite{Sch35b}

The Cat Experiment was supposed to support the out-of-focus photograph
option, the view that the Uncertainty Principle is just a restriction on
our \emph{knowledge} of reality. Schr\"odinger thought it was obvious that
the cat couldn't actually be somehow neither alive nor dead. Like
Einstein, Schr\"odinger favoured the view that the quantum description is
incomplete, and that reality contains further details, hidden behind
Heisenberg's veil.

In the decades since 1935, most physicists who care about these issues
have concluded that Einstein and Schr\"odinger were wrong. Bell's Theorem,
together with the quantum predictions being confirmed by Clauser,
Aspect, Zeilinger and many others, has often been interpreted as showing
that the spooky action at a distance which Einstein hoped to avoid with
additional `hidden variables', is an inevitable part of the quantum
world.

Retrocausality is already the most interesting challenge to that view.
By taking the first option on Schr\"odinger's list -- by treating QM as an
unavoidably fuzzy picture of a sharper reality -- it can allow the kind
of quantum black boxes needed to avoid retrosignalling and paradoxes.
How satisfying, then, if it also explains the other thing that
Schr\"odinger put his finger on in 1935, when he invented the term
`entanglement', and called it `\emph{the} characteristic trait of
quantum mechanics.'\footnote{We are grateful to Emily Adlam, Guido Bacciagaluppi and Nathan Argaman for comments
  on previous versions. HP is grateful to audiences in Toronto, Bonn and Dortmund.}

\end{document}